# Unusually Strong Lateral Interaction in the CO overlayer in Phosphorene-based systems

A. Politano[1](✉), M.S. Vitiello[2], L. Viti[2], J. Hu[3], Z.Q. Mao[3], J. Wei[3], G. Chiarello[1] and D. W. Boukhvalov[4,5]

*1 Department of Physics, University of Calabria, via ponte Bucci, 31/C, 87036 Rende (CS), Italy*

*2 NEST, Istituto Nanoscienze–CNR and Scuola Normale Superiore Piazza San Silvestro 12 , Pisa I-56127 , Italy*

*3 Department of Physics and Engineering Physics, Tulane University, New Orleans LA 70118 USA*

*5 Department of Chemistry, Hanyang University, 17 Haengdang-dong, Seongdong-gu, Seoul 133-791, Republic of Korea*

*[6]Theoretical Physics and Applied Mathematics Department, Ural Federal University, Mira Street 19, 620002 Ekaterinburg, Russia*

## ABSTRACT

By means of vibrational spectroscopy and density functional theory (DFT), we investigate CO adsorption on phosphorene-based systems. We find stable CO adsorption at room temperature on both phosphorene and bulk black phosphorus. The adsorption energy and the vibrational spectrum have been calculated for several possible configurations of the CO overlayer. We find that the vibrational spectrum is characterized by two different C-O stretching energies. The experimental data are in good agreement with the prediction of the DFT model and unveil the unusual C-O vibrational band at 165-180 meV, activated by the lateral interactions in the CO overlayer.

## 1 Introduction

Recently, black phosphorus (BP) has attracted great interest[1] since the presence of a narrow band gap makes it a more suitable candidate compared to graphene for an abundant number of device applications (flexible electronics [2], nanoresonators [3], Terahertz photodetection [4], charge trap memories[5], AM demodulators [2]). A single BP layer is named phosphorene[6]. The bulk crystal of BP has a phosphorene surface termination. Thus, the surface of BP is the archetype of a supported phosphorene system.

The chemical reactivity of free-standing phosphorene and bulk crystals of BP deserves particular attention for understanding the pitfalls of phosphorene-based devices arising from chemical stability in ambient conditions[7-9] and, moreover, for exploring its possible use in catalysis[10] and as gas sensor [11, 12].

While oxidation processes have been recently explored [9, 13-15], the reactivity of phosphorene toward CO is hitherto unclear. First-principle calculations found that CO weakly interacts with phosphorene [16], even if no experimental studies have been carried out yet. Several possible reactions would be possible[17], if CO could be stably adsorbed on phosphorene-based systems.

High-resolution electron energy loss spectroscopy (HREELS) is an effective tool for investigating adsorption at solid surfaces, also in consideration of its versatility for both vibrational and electronic excitations. From the analysis of the C-O intramolecular vibration, it is possible to infer information on CO-substrate interaction, its related charge transfer and, moreover, on adsorption sites [18].

Herein, we report on CO interaction with phosphorene-based systems at room temperature by means of HREELS experiments and density functional theory (DFT) calculations. Contrarily to previous reports, we here provide evidence of stable CO adsorption at room temperature. We demonstrate that CO adsorption occurs with the formation of pairs and rows, whose vibrational spectrum is characterized by an unusually weak C-O intramolecular bond. In contrast with the case of CO adsorption on metal surfaces, the interaction between CO molecules already starts with CO pairs. The total binding energy per molecule remains almost unmodified with coverage.

## 2 Experimental methods

The BP single crystal was synthesized using a chemical vapor transport method. A mixture of red phosphorus, AuSn, and SnI$_4$ powder with mole ratio 1000:100:1 was sealed into an evacuated quartz tube. The tube is then placed into a double-zone tube furnace with temperature set at 870 K and 770 K for the hot and cold end, respectively. Large single crystals can be obtained after one week of transport. Grown BP samples exhibit ambipolar behavior, a gate-dependent metal–insulator transition, and mobility up to 4000 cm$^2$ V$^{-1}$ s$^{-1}$, as reported elsewhere [7]. Figure 1a shows a selected-area electron diffraction (SAED) pattern of our grown samples, acquired with transmission electron microscopy at Tulane University, USA. Exfoliated BP samples show atomic flatness and large-scale terraces, as evidenced by atomic force microscopy (AFM) experiments, reported in Figure 1b.

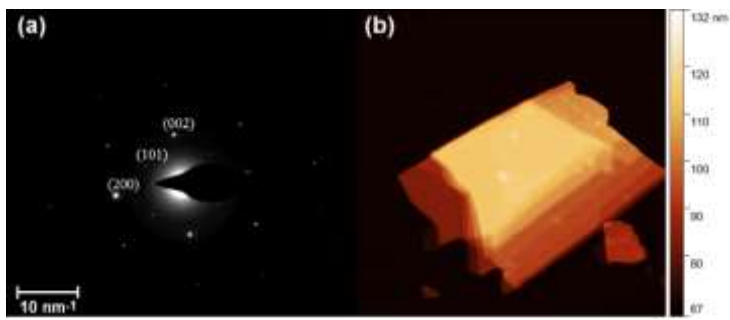

**Figure 1.** (a) SAED pattern of BP. Panel (b) shows an AFM image, which puts in evidence the existence of large-scale terraces in exfoliated BP samples.

HREELS experiments have been carried out in an ultra-high vacuum (UHV) chamber operating at a base pressure of 5·10$^{-9}$ Pa at University of Calabria, Italy. The sample has been exfoliated in situ via micromechanical cleavage under UHV conditions. Surface cleanliness and structural order have been assessed via a combination of X-ray photoelectron spectroscopy (XPS) and Auger electron spectroscopy (AES) and low-energy electron diffraction (LEED) measurements, respectively. HREELS spectra have been measured by using an electron energy loss spectrometer (Delta 0.5, SPECS) with an angular acceptance of ±0.5°. The energy resolution of the spectrometer has been degraded to 5 meV, so as to increase the signal-to-noise ratio of loss peaks. Both CO exposure and HREELS experiments have been performed at room temperature.

## 3 Results and discussion

The atomic structure and energetics of various configurations of CO on phosphorene have been studied by DFT using the QUANTUM-ESPRESSO code [19] and the GGA–PBE + van der Waals (vdW) approximation, feasible for the description of the adsorption of molecules on surfaces [20, 21] with the employment of ultrasoft pseudopotentials [22]. We used energy cutoffs of 25 Ry and 400 Ry for the plane-wave expansion of the wave functions and the charge density, respectively, and the 6×6×1 Monkhorst-Pack $k$-point grid for the Brillouin sampling [23]. For the modeling of the surface of BP, we used a rectangular supercell of monolayer phosphorene containing 72 phosphorus atoms (see Figure 2), previously employed for studying the covalent functionalization of phosphorene[24]. For the case of adsorption of a single CO molecule on phosphorene, we performed calculations with optimization of both the atomic positions and lattice parameters. We find deviation from standard values (a = 3.34 Å, b = 4.57 Å) of less than 0.8%. Based on this finding, in further modeling we omitted the optimization of lattice parameters. Adsorption energies were calculated by the standard formula:

$E_{ads} = [(E_{pure\ phosphorene} + nE_{CO}) - E_{phosphorene+nCO}]/n$.

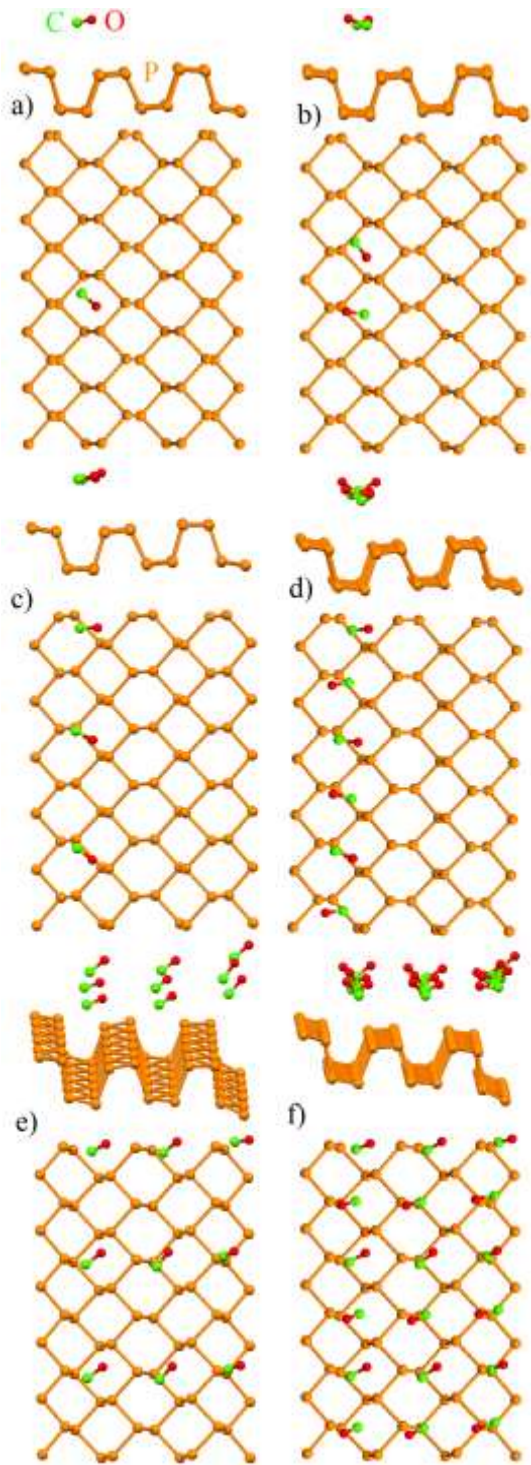

**Figure 2** Side and top views of optimized atomic structure of single (a) and pair (b) of CO molecules on phosphorene. Panel (c) and (d) represent CO rows aligned along one direction of phosphorene. In panels (e) and (f) two different types of uniform coverage are reported.

To estimate the energy of the CO-CO interaction, we take the optimized atomic structure of the investigated configuration. The difference between the total energy per adsorbed CO molecule and the free-standing CO molecule is defined as the energy of CO-CO interaction. Vibrational frequencies were calculated by stretching CO molecules along the carbon-oxygen bond by 0.04 Å. Energy difference was also calculated in the upstretched case. Various amounts of CO on phosphorene from single molecules to full coverage were modelled (Figure 2, panels a-f). The adsorption energies and the CO-CO interaction energies for each configuration are reported in Table I where the distance of CO molecules from the surface plane is also given.

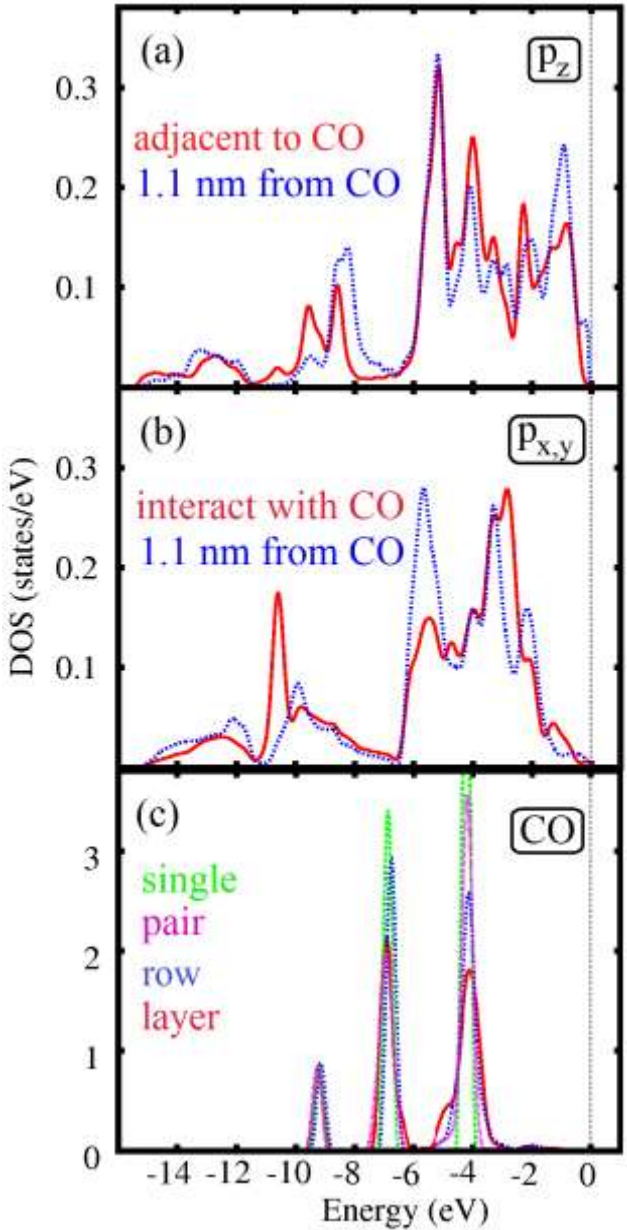

**Figure 3** PDOS of (a,b) 3p orbitals of phosphorus atoms adjacent to single CO molecule and 1.1 nm far from adsorbed CO molecule, and (c) CO in various adsorption configurations (single – Fig. 2a, pair – Fig. 2b, row – Fig. 2d and full monolayer, denoted as "layer" – Fig. 2f).

Similarly to CO adsorption on metal substrates [18], in the case of CO/phosphorene different possible adsorption sites over phosphorus atoms of the surface exist: on-top of a single substrate atom, bridge between two atoms of substrate and over the center of the void between atoms (i.e. top, bridge and hole positions). We have examined all these configurations, finding that adsorption over the hole (Figure 2a) is more energetically favorable (by more than 20 meV) compared with the on-top adsorption site. Conversely, CO adsorption in the bridge site is unstable. Interestingly, we note that in contrast to adsorption on metal substrates[18], the axis of adsorbed CO molecules significantly deviates from the surface normal. This can be explained by considering that, for the case of CO/phosphorene, carbon orbitals are bound with lone pairs of 3p orbitals of phosphorus, which is located above and below of phosphorene plane (see detailed discussion in Ref. [24]), while in the case of CO adsorption on metal surfaces, the d bands of the underlying metal substrate have a key role in the formation of the CO-metal bond [25]. The adsorption of CO molecules influences the electronic structure of $p_z$ orbitals (lone pairs) around the Fermi level (Fig. 3a). In addition, CO adsorption modifies the electronic structure of $p_z$ and $p_y$ orbitals (Fig. 3b) participating in the formation of σ-bonds within the phosphorene layer.

**Table I** Values of the CO adsorption energy and the CO-CO interaction energy for the various configurations reported in Figure 2. The distance of CO molecules from the surface plane is also reported.

| Configuration | $E_{ads}$ (meV) | $E_{CO-CO}$ (meV) | $d_{CO-plane}$ (nm) | Vibrational energy (meV) |
|---|---|---|---|---|
| A | 106.2 | | 0.269 | 259 |
| B | 106.7 | 7.6 | 0.252 | 220 and |

|   |       |      |       | 165         |
|---|-------|------|-------|-------------|
| C | 72.6  | 1    | 0.271 | 220         |
| D | 104.4 | 2.4  | 0.270 | 220 and 165 |
| E | 88.6  | 15.3 | 0.268 | 165-168     |
| F | 99.3  | 22.8 | 0.265 | 162-170     |

The electronic structure of a single CO molecule adsorbed on phosphorene (Fig. 3c) is close to that of free-standing CO molecules. Correspondingly, the C-O vibrational energy for single CO molecule is 259 meV, i.e. an energy similar to that of CO in the gas phase (266.7 meV [26]). For evaluating the role of defects, such as grain-boundaries, we performed calculations of CO adsorption in vicinity of Stone-Wales defects in phosphorene monolayer. Obtained values of the CO-plane distance, the binding energy and the vibrational energy are almost the same as in the case of pristine phosphorene (0.266 nm, 108.4 meV and 262 meV, respectively). Thus, we can affirm that the presence of defects in phosphorene plays insignificant role in CO adsorption.

As next step, we modeled the adsorption of two CO molecules and examined different configurations of pairs. We find that the more energetically favorable configuration of the CO pair is the one shown on Figure 2b. The random adsorption of CO pairs is less favorable by 11-20 meV, depending on the configuration. The C-O vibrational band in CO pairs is composed of two normal modes. The first component arises from the mutual interaction between CO molecules, which shifts C-O energy from 259 to 220 meV. The adsorption of one CO molecule onto a site, where it can interact with lone pairs of two P atoms, implies a stronger CO-phosphorene bond, with a consequent softening of the intramolecular C-O bond. Therefore, the C-O vibrational energy of CO pairs has a band with lower frequencies. Interestingly, the C-O vibrational energy softens down to 165 meV, pointing to an unusually strong CO-substrate interaction. This induces changes in both 3p orbitals of P participating in the P-CO bond and in P-P bonds (Fig. 3a,b), which are responsible of the lateral interaction between CO molecules adsorbed on phosphorene (Fig. 3c). The significant influence of chemisorption of adsorbed species on the electronic structure of phosphorene was previously discussed in several papers reporting the superb capabilities of phosphorene for gas detection[11, 12, 27].

Such a low C-O vibrational energy deserves particular attention. For strongly bonded CO molecules in four-fold adsorption sites on metals, the C-O energy is ~210 meV [18]. Values as low as ~180 meV have been reported for complexes formed by CO molecules and alkali atoms [28]. The absence of noticeable changes in the distance between CO molecules and the phosphorene plane ($d_{CO\text{-plane}}$) and, moreover, in adsorption energies (see Table I) indicates that the softening of vibrational modes is not related to modifications in the atomic structure of CO-phosphorene bonds.

We suggest that the occurrence of a C-O mode at 165 meV in CO/phosphorene indicates the population of antibonding orbitals of CO molecules, activated by the charge transfer with the bulk BP. For evaluating the role of CO-CO coupling, the most energetically favorable pair configuration along one of the axis of phosphorene (Figure 2d) is taken as unit to form a CO row and then one of "sublattices" of CO is removed from the row (Figure 2c). Calculations of vibrational frequencies indicate the presence of a single mode at 220 meV, which is accompanied by another vibration at 165 meV only in the case of dense adsorption of CO molecules (Figure 2d), i.e. whenever the CO-CO interaction energy becomes not negligible or, likewise, whenever the CO-CO distance is sufficiently low.

As a final step, we studied the effects of the full CO coverage of the phosphorene substrate, obtained by multiplying the row of CO molecules shown on

Figure 2d. For the case of the most dense CO coverage (Figure 2f), we obtain vibrational modes at 162-170 meV, corresponding to the largest values of CO-CO binding energies. Removal of one "sublattice" from the dense structure of panel (f) gives the configuration in panel (e), which is characterized by almost the same values of vibrational energies and rather high energies of CO-CO interaction (see Table I). Strikingly, the optimized orientation of CO molecules in the configuration (e) is the same of that in panel (f), i.e. the molecular orientation survives regardless the removal of one CO sublattice.

To check the validity of the employment of monolayer phosphorene as a model system for the surface of BP, we performed additional calculations for adsorption of single CO molecule on bi- and trilayer phosphorene. Negligible increasing of binding energy between CO and phosphorene ($E_{ads}$=112 and 114 meV respectively) was found on bilayer phosphorene. Thus, we can conclude that monolayer phosphorene is a quite feasible model for describing adsorption on the surface of BP. However, it should be noticed that the value of the vibrational frequency of C-O intramolecular stretching of a single CO molecule on bi- and trilayer phosphorene decreased to 248 and 245 meV respectively. Therefore, it is expected that single CO molecule on bulk BP could have even inferior C-O stretching energy.

To validate theoretical results, HREELS experiments on CO adsorbed on BP at room temperature have been performed. The BP sample was grown and characterized by means of experimental methods described in details in the Electronic Supplementary Information.

Figure 4a shows the vibrational data of a BP sample saturated with CO at room temperature. The CO coverage has been estimated by a calibration procedure using X-ray photoelectron spectroscopy and Auger electron spectroscopy to be ~0.12 ML (1 ML here is defined as the ratio between the number of adsorbed species and the number of the atoms in the outermost surface layer).

The vibrational spectrum of the clean surface is featureless, while CO exposure induces the appearance of a broad band arising from C-O intramolecular stretching vibrations. After the subtraction of an exponential background (Figure 4b), the resulting spectra have been fitted with two Gaussian line-shapes peaked at 178 and 235 meV. The surface is saturated with CO for doses around 10 L and no variation of the vibrational spectrum has been recorded even upon large CO doses.

The presence of a splitting should be related to the existence of CO rows in the CO overstructure. The vibrational band corresponding to the structure in Figure 2d well reproduces the experimental result. We point out that in UHV conditions and at room temperature the CO coverage does not exceed ~0.12 ML. Thus, we cannot reproduce the configuration in panels (e) and (f) of Figure 2 and, as a consequence, we do not observe a single C-O peak at 162-170 meV, which characterizes the vibrational spectrum of a full monolayer of CO on phosphorene (Figure 2f).

Plenty of non-equivalent adsorption sites exist for CO on phosphorene surfaces. Thus, CO molecules on phosphorene do not self-organize in ordered patterns. For this reason, the experimental vibrational band is very broad. The adsorption of the first CO molecule significantly changes the binding energy between different sites and neighbor molecules: slightly different binding energies imply slightly different C-O bond strength and, as result, different frequencies.

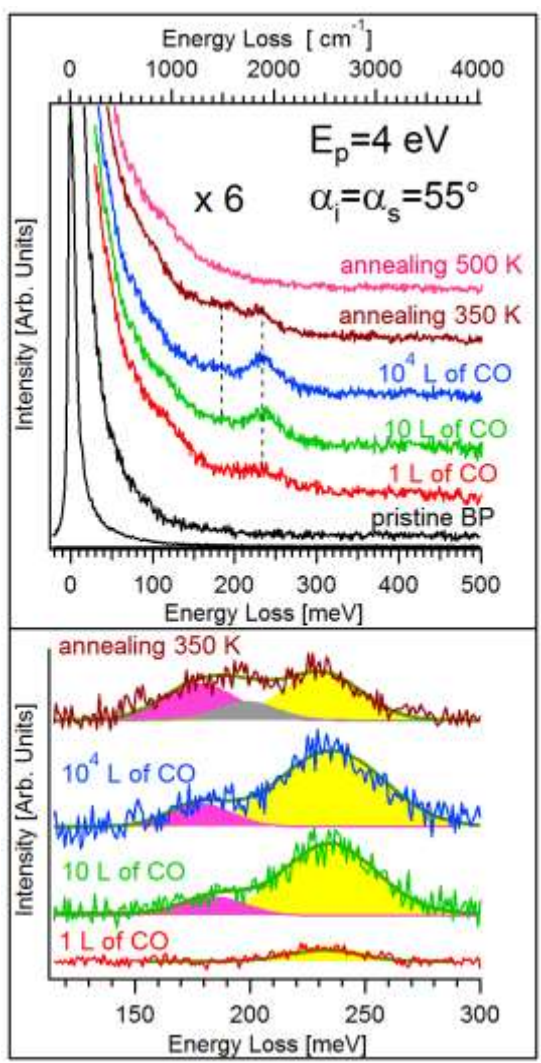

**Figure 4** (a) HREELS spectra for clean BP and the CO-saturated BP surface, successively annealed at 350 and 500 K. The sample has been exposed to 1, 10 and $10^4$ L of CO (1 L=1 torr·s) at room temperature. The top and the bottom axis report values of vibrational energies in cm$^{-1}$ and meV, respectively. (b) After background subtraction with a polynomial function, the resulting vibrational spectra have been fitted by Gaussian line-shapes.

Upon annealing at 350 K, the area of the whole region of the C-O intramolecular stretching decreased by 21%, thus indicating the partial desorption of CO molecules induced by sample heating. In particular, the area of the C-O mode at higher energy decreases by 53% and its frequency red-shifted by 4 meV. Conversely, the area of the peak at 178 meV increases by 82%, pointing to a heating-induced rearrangement of CO rows. Annealing also induces the emergence of a feature at 199 meV.

Further heating at 500 K induces the disappearance of all spectral features in the band of C-O intramolecular stretching, therefore indicating the total desorption of CO molecules from the BP surface.

Previously, it has been reported that CO adsorption on phosphorene induces a 0.03 $e$ charge transfer with an adsorption strength of 0.325 eV/unit cell [12]. A weak binding energy of CO with phosphorene has been reported in Ref. [12]. By contrast, our theoretical and experimental results support stable adsorption of CO on phosphorene.

Recently, scanning tunneling microscopy (STM) has been used for imaging a hexagonal adlayer with $P_{13}$ clusters with (4√3 × 4√3)–R30° symmetry grown on Pt(111) [29]. Therein [29], it has been reported that the hexagonal phosphorus adlayer is quite unreactive toward CO. While it is quite expected that the sticking coefficient of CO on the $P_{13}$ clusters on Pt(111) could be reduced with respect to transition metals, our results also disagree with findings in Ref. [29]. However, the chemical reactivity of hexagonal $P_{13}$ clusters could be different with respect to that of phosphorene. The weak CO adsorption on $P_{13}$ clusters is mainly determined by both the geometric structure and the orbital ordering, while the presence of the underlying metal substrate plays minor role. In the case of free-standing phosphorene, four orbitals with lone pairs are almost crossing over the hole region. The interaction of eight electrons in the four lone pairs of phosphorene with one CO molecule provides rather high value of CO binding energies, with subsequent softening of the C-O bond. In the case of a hexagonal phosphorus overlayer supported by a metal substrate, CO interacts with 4-6 electrons of phosphorene orbitals (2 spins of xz, yz, and $z^2$ orbitals). As a result, the binding energy is smaller by about 1.5 times and also the softening of the C-O intramolecular bond is reduced compared with the case of free-standing phosphorene. In the case of $P_{13}$, three $P_4$ tetrahedral clusters with one P in the center exist. All orbitals with lone pairs are strictly oriented along the line center of the P tetrahedron. Thus, CO could interact only with one orbital (only 2 electrons) with consequently weak $P_{13}$-CO bonding.

Finally, we can suggest that the strong bonding between CO and phosphorene can be used also for CO-induced chemical scission of phosphorene nanoribbons [30].

# 4 Conclusions

We have demonstrated that CO stably adsorbs on phosphorene at room temperature. Our results indicate that different CO overstructures coexist on the phosphorene surface. Configurations with similar adsorption energies are characterized by different vibrational spectra, with the emergence of an unusual band at 165-180 meV, activated in conditions enhancing lateral interactions between CO adsorbed molecules. Our results pave the way for phosphorene-based catalysis and for engineering molecularly assembled capping layers on the phosphorene surface for its protection from oxidation.

**Acknowledgments**

The work at Tulane is supported by the US Department of Energy under grant DE-SC0014208 (support for single-crystal growth and structure characterization).

**References**

[1] Wang, Y.;Cong, C.;Fei, R.;Yang, W.;Chen, Y.;Cao, B.;Yang, L.; Yu, T. Remarkable anisotropic phonon response in uniaxially strained few-layer black phosphorus. *Nano Res.* **2015**, 10.1007/s12274-015-0895-7, 1-10.

[2] Zhu, W.;Yogeesh, M. N.;Yang, S.;Aldave, S. H.;Kim, J.-S.;Sonde, S.;Tao, L.;Lu, N.; Akinwande, D. Flexible Black Phosphorus Ambipolar Transistors, Circuits and AM Demodulator. *Nano Lett.* **2015**, *15*, 1883-1890.

[3] Wang, C.-X.;Zhang, C.;Jiang, J.-W.;Park, H. S.; Rabczuk, T. Mechanical strain effects on black phosphorus nanoresonators. *Nanoscale* **2016**, *8*, 901-905.

[4] Viti, L.;Hu, J.;Coquillat, D.;Knap, W.;Tredicucci, A.;Politano, A.; Vitiello, M. S. Black Phosphorus Terahertz Photodetectors. *Adv. Mater.* **2015**, *27*, 5567–5572.

[5] Feng, Q.;Yan, F.;Luo, W.; Wang, K. Charge trap memory based on few-layer black phosphorus. *Nanoscale* **2016**, *8*, 2686-2692.

[6] Zhang, X.;Xie, H.;Liu, Z.;Tan, C.;Luo, Z.;Li, H.;Lin, J.;Sun, L.;Chen, W.;Xu, Z.;Xie, L.;Huang, W.; Zhang, H. Black Phosphorus Quantum Dots. *Angew. Chem.* **2015**, *54*, 3653–3657.

[7] Nathaniel, G.;Darshana, W.;Yanmeng, S.;Tim, E.;Jiawei, Y.;Jin, H.;Jiang, W.;Xue, L.;Zhiqiang, M.;Kenji, W.;Takashi, T.;Marc, B.;Yafis, B.;Roger, K. L.; Chun Ning, L. Gate tunable quantum oscillations in air-stable and high mobility few-layer phosphorene heterostructures. *2D Materials* **2015**, *2*, 011001.

[8] Island, J. O.;Steele, G. A.;van der Zant, H. S. J.; Castellanos-Gomez, A. Environmental instability of few-layer black phosphorus. *2D Materials* **2015**, *2*, 011002.

[9] Wood, J. D.;Wells, S. A.;Jariwala, D.;Chen, K.-S.;Cho, E.;Sangwan, V. K.;Liu, X.;Lauhon, L. J.;Marks, T. J.; Hersam, M. C. Effective Passivation of Exfoliated Black Phosphorus Transistors against Ambient Degradation. *Nano*

*Lett.* **2014**, *14*, 6964–6970.

[10] Sa, B.;Li, Y.-L.;Qi, J.;Ahuja, R.; Sun, Z. Strain Engineering for Phosphorene: The Potential Application as a Photocatalyst. *J. Phys. Chem. C* **2014**, *118*, 26560-26568.

[11] Ray, S. J. First-principles study of $MoS_2$, phosphorene and graphene based single electron transistor for gas sensing applications. *Sens. Actuators B: Chem.* **2016**, *222*, 492-498.

[12] Kou, L.;Frauenheim, T.; Chen, C. Phosphorene as a superior gas sensor: Selective adsorption and distinct I-V response. *J. Phys. Chem. Lett.* **2014**, *5*, 2675-2681.

[13] Wang, G.;Pandey, R.; Karna, S. Phosphorene Oxide: Stability and electronic properties of a novel 2D material. *Nanoscale* **2014**, *7*, 524-531.

[14] Edmonds, M. T.;Tadich, A.;Carvalho, A.;Ziletti, A.;O'Donnell, K. M.;Koenig, S. P.;Coker, D. F.;Özyilmaz, B.;Neto, A. H. C.; Fuhrer, M. S. Creating a Stable Oxide at the Surface of Black Phosphorus. *ACS Appl. Mater. Interfaces* **2015**, *7*, 14557–14562.

[15] Favron, A.;Gaufres, E.;Fossard, F.;Phaneuf-Lheureux, A.-L.;Tang, N. Y. W.;Levesque, P. L.;Loiseau, A.;Leonelli, R.;Francoeur, S.; Martel, R. Photooxidation and quantum confinement effects in exfoliated black phosphorus. *Nat. Mater.* **2015**, *14*, 826-832.

[16] Cai, Y.;Ke, Q.;Zhang, G.; Zhang, Y.-W. Energetics, Charge Transfer, and Magnetism of Small Molecules Physisorbed on Phosphorene. *J. Phys. Chem. C* **2015**, *119*, 3102-3110.

[17] Puschmann, F. F.;Stein, D.;Heift, D.;Hendriksen, C.;Gal, Z. A.;Grützmacher, H.-F.; Grützmacher, H. Phosphination of Carbon Monoxide: A Simple Synthesis of Sodium Phosphaethynolate (NaOCP). *Angew. Chem.* **2011**, *50*, 8420-8423.

[18] Politano, A.; Chiarello, G. Vibrational investigation of catalyst surfaces: change of the adsorption site of CO molecules upon coadsorption. *J. Phys. Chem. C* **2011**, *115*, 13541-13553.

[19] Giannozzi, P.;Baroni, S.;Bonini, N.;Calandra, M.;Car, R.;Cavazzoni, C.;Ceresoli, D.;Chiarotti, G. L.;Cococcioni, M.;Dabo, I.;Dal Corso, A.;de Gironcoli, S.;Fabris, S.;Fratesi, G.;Gebauer, R.;Gerstmann, U.;Gougoussis, C.;Kokalj, A.;Michele, L.;Martin-Samos, L.;Marzari, N.;Mauri, F.;Mazzarello, R.;Paolini, S.;Pasquarello, A.;Paulatto, L.;Sbraccia, C.;Scandolo, S.;Sclauzero, G.;Seitsonen, A. P.;Smogunov, A.;Umari, P.; Wentzcovitch, R. M. QUANTUM ESPRESSO: a modular and open-source software project for quantum simulations of materials. *J. Phys.: Condens. Matter* **2009**, *21*, 395502.

[20] Perdew, J. P.;Burke, K.; Ernzerhof, M. Generalized Gradient Approximation Made Simple. *Phys. Rev. Lett.* **1996**, *77*, 3865-3868.

[21] Barone, V.;Casarin, M.;Forrer, D.;Pavone, M.;Sambi, M.; Vittadini, A. Role and effective treatment of dispersive forces in materials: Polyethylene and graphite crystals as test cases. *J. Comput. Chem.* **2009**, *30*, 934-939.

[22] Vanderbilt, D. Soft self-consistent pseudopotentials in a generalized eigenvalue formalism. *Phys. Rev. B* **1990**, *41*, 7892-7895.

[23] Monkhorst, H. J.; Pack, J. D. Special points for Brillouin-zone integrations. *Phys. Rev. B* **1976**, *13*, 5188-5192.

[24] Boukhvalov, D. W.;Rudenko, A. N.;Prishchenko, D. A.; Mazurenko, V. G.; Katsnelson, M. I. Chemical modifications and stability of phosphorene with impurities: a first principles study. *Phys. Chem. Chem. Phys.* **2015**, *17*, 15209-15217.

[25] Stroppa, A.;Termentzidis, K.;Paier, J.;Kresse, G.; Hafner, J. CO adsorption on metal surfaces: A hybrid functional study with plane-wave basis set. *Phys. Rev. B* **2007**, *76*, 195440.

[26] Sterrer, M.;Yulikov, M.;Risse, T.;Freund, H.-J.;Carrasco, J.;Illas, F.;Di Valentin, C.;Giordano, L.; Pacchioni, G. When the Reporter Induces the Effect: Unusual IR spectra

of CO on $Au_1$/MgO(001)/Mo(001). *Angew. Chem.* **2006**, *45*, 2633-2635.

[27] Cui, S.;Pu, H.;Wells, S. A.;Wen, Z.;Mao, S.;Chang, J.;Hersam, M. C.; Chen, J. Ultrahigh sensitivity and layer-dependent sensing performance of phosphorene-based gas sensors. *Nat. Commun.* **2015**, *6*, 8632.

[28] Politano, A.;Chiarello, G.;Benedek, G.;Chulkov, E. V.; Echenique, P. M. Vibrational measurements on alkali coadsorption systems: experiments and theory. *Surf. Sci. Rep.* **2013**, *68*, 305–389.

[29] Heikkinen, O.;Pinto, H.;Sinha, G.;Hämäläinen, S. K.;Sainio, J.;Öberg, S.;Briddon, P. R.;Foster, A. S.; Lahtinen, J. Characterization of a Hexagonal Phosphorus Adlayer on Platinum (111). *J. Phys. Chem. C* **2015**, *119*, 12291-12297.

[30] Xihong, P.; Qun, W. Chemical scissors cut phosphorene nanostructures. *Mater. Res. Express* **2014**, *1*, 045041.